\documentclass[reprint,amsmath,amssymb,pra,aps, nofootinbib]{revtex4-1}

\usepackage[T1]{fontenc}
\usepackage[utf8]{inputenc}
\usepackage{amsmath, amssymb}
\usepackage{bbold}
\usepackage{bm}
\usepackage{tipa}
\usepackage{blindtext}
\usepackage[dvipsnames]{xcolor}
\definecolor{dcyan}{RGB}{0,100,100}
\usepackage{graphicx}
\usepackage{tabularx}

\newcommand{\bra}[1]{\langle{#1}|}
\newcommand{\ket}[1]{|{#1}\rangle}

\newcommand{\Fclr}{blue}
\newcommand{\Bclr}{red}

\begin{document}
\title{Understanding and suppressing backscatter in optical resonators}
\author{Matt Jaffe}
\author{Lukas Palm}
\author{Claire Baum}
\author{Lavanya Taneja}
\author{Aishwarya Kumar}
\author{Jonathan Simon}

\affiliation{The Department of Physics, The James Franck Institute, and The Pritzker School of Molecular Engineering, The University of Chicago, Chicago, IL}

\date{\today}

\begin{abstract}
   Optical cavities have found widespread use in interfacing to quantum emitters. Concerns about backreflection and resulting loss, however, have largely prevented the placement of optics such as lenses or modulators within high-finesse cavities. In this work, we demonstrate a million-fold suppression of backreflections from lenses within a twisted optical cavity. We achieve this by quantitatively exploring backscatter in Fabry-P\'erot resonators, separating the effect into three physical sectors: polarization, mode envelope and spatial mode profile. We describe the impact of each of these sectors, and demonstrate how to minimize backreflections within each. This culminates in measured effective reflectivities below the part-per-billion level for the fundamental mode. Additionally, we show that beams carrying orbital angular momentum experience up to $10^{4}$ times additional suppression, limited only by the density of states of other cavity modes. Applying these ideas to laser gyroscopes could strongly suppress lock-in, thereby improving sensitivity at low rotation rates.
\end{abstract}

\maketitle

\section{Introduction}

High-finesse optical cavities are broadly useful across quantum optics, enabling efficient interfaces of individual photons to quantum emitters such as atoms and ions~\cite{walther2006cavity}, quantum dots~\cite{Yoshle2004, Reithmaier2004}, rare-earth ions~\cite{zhong2018optically,raha2020optical}, and defect centers ~\cite{faraon2012coupling,riedrich2014deterministic}. Intracavity optics, such as lenses or modulators, could radically expand the capabilities of such optical resonators. While refractive elements are used extensively in free space optics, they have thus far remained largely absent from moderate- to high- finesse optical cavities. Concerns about backscatter and loss have typically been presumed to preclude intracavity changes of refractive index, even if anti-reflection (AR) coatings are employed. If these concerns could be addressed, intracavity optics would be transformative, enabling new capabilities for light-matter interaction.

Backscattering has been studied across the electromagnetic spectrum, from microwave to optical frequencies. It is of practical importance for applications in ring lasers~\cite{Menegozzi1973, Haus1985} and gyroscopes~\cite{Chow1985, Etrich1992}, and is fundamentally connected to topological systems~\cite{lodahl2017chiral}. In particular, backscatter immunity is a defining characteristic of topologically protected edge channels, even in the presence of disorder~\cite{Wang2009,Hafezi2011,hafezi2013imaging,Zhang2021}. Elimination of undesired optical backreflections, however, has so far been limited to active~\cite{Krenz2007} or passive~\cite{svela2020coherent} cancellation.

In this work, we identify three main sectors contributing the the total amount of backreflection: polarization, mode envelope and spatial mode profile. We experimentally and quantitatively explore the effects of each of these sectors, and minimize the total backreflection by optimizing over each. These approaches amount to either suppression of the back-coupling matrix element or reduction of the available density of states for backscattering. We demonstrate that backscattering can be suppressed nearly one million-fold in a twisted optical cavity, allowing for effective reflectivities below 1 part per billion (ppb). We find that engineering the polarization eigenstates of the cavity is the essential tool for achieving this performance. Finally, we show that beams carrying orbital angular momentum (OAM) exhibit even stronger suppression of backreflections due to a topological protection arising from their phase winding.

\section{Reflector in a cavity}

\begin{figure*} 
	\centering
	\includegraphics[width=0.98\textwidth]{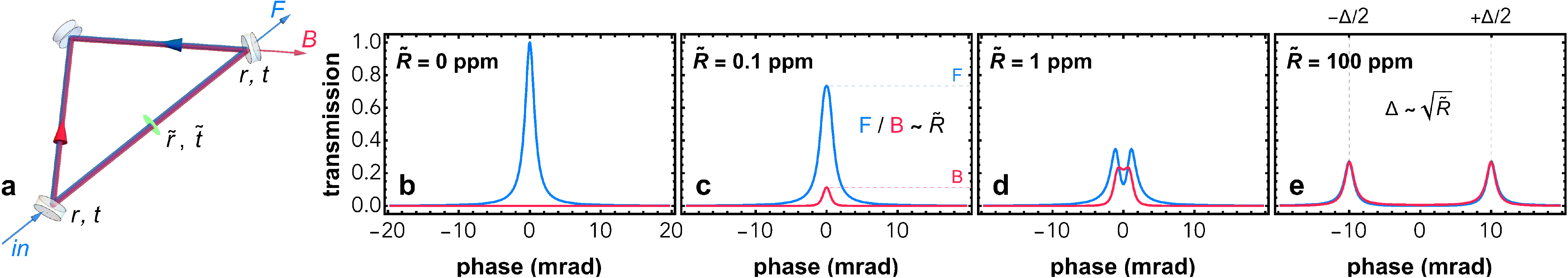}
	\caption{
		\textbf{A simple model of intracavity backscattering}. (a) We consider a planar running wave cavity with a single intracavity reflector. Input and output mirrors have amplitude reflection (transmission) coefficient $r$ ($t$). The third mirror is a perfect reflector. The intracavity reflector (green) has amplitude reflection (transmission) coefficient $\tilde{r}$ ($\tilde{t}$). Output paths of the forward- and backward- modes from the input mirror are omitted. (b-e) Cavity transmission for different values of $\tilde{R} = |\tilde{r}|^{2}$. For $\tilde{R} = 0$, we see the expected Lorentzian lineshape. The cavity finesse has been set to 4000  via $r$ and $t$ with no loss. The \Fclr{} (\Bclr{}) curve shows the forward-traveling mode $F$ (backward-traveling mode $B$) transmission. As $\tilde{R}$ increases, we see coupling of the injected forward-traveling light into the backward-traveling mode. For sufficiently small $\tilde{R}$, the forward and backward responses are both approximately Lorentzian with peak ratio $\propto\tilde{R}$. For larger $\tilde{R}$, the forward and backward modes hybridize, splitting into two spectrally-separated peaks with near-equal participation of both $F$ and $B$. The splitting of these peaks $\Delta$~is~$\propto\sqrt{\tilde{R}} = \left|\tilde{r}\right|$, and is directly analogous to the classical coupling of two harmonic oscillators.
	}
		\label{fig:reflector_in_a_cavity}
\end{figure*}

To begin, we consider a running wave optical cavity consisting of three mirrors as shown in Fig.~\ref{fig:reflector_in_a_cavity}a. The two input/output mirrors have amplitude reflection (transmission) coefficients $r$ ($t$), while the third mirror is a perfect reflector. An intracavity reflector with amplitude reflection (transmission) coefficient $\tilde{r}$ ($\tilde{t}$) can couple a forward-traveling mode to a backward-traveling mode. Indeed, even in cavities without an explicit reflector, the mirror imperfections in high-finesse ring cavities can be sufficient to induce detrimental backscattering~\cite{Krenz2007}.

Using standard input/output field relations~\cite{siegman86}, the intracavity and output fields for both the forward- and backward-traveling modes may be computed, as shown in Figs.~\ref{fig:reflector_in_a_cavity}(b-e) for various values of $\tilde{R} = |\tilde{r}|^{2}$. As the intracavity reflectivity $\tilde{R}$ increases, the forward and backward modes hybridize until they are fully mixed. Note that the highest reflectivity shown, $\tilde{R} = 100$ parts per million (ppm), corresponds approximately to the  best (lowest-reflectivity) commercially-available AR coatings. Thus, even at the modest finesse of 4000 shown in Figs.~\ref{fig:reflector_in_a_cavity}(b-e), any optic within the cavity would fully hybridize the forward and backward modes. Note that unlike the case of a standing wave cavity~\cite{Jayich2008}, sub-wavelength changes of the single reflector's longitudinal position do not impact the amount of backreflection for the running wave case.

A powerful tool to break the forward-backward symmetry is a non-planar cavity geometry, which provides a round-trip image/polarization rotation~\cite{Nilsson1989, Yelland1994} (an example of a Pancharatnam-Berry phase~\cite{Pancharatnam1956, Berry1984}). Combined with the Faraday effect, this rotation allows introduction of loss for one polarization state via Brewster reflection and thus unidirectional lasing of an active gain medium within such a cavity~\cite{Kane1985, Maker1993}. For a resonator without gain, this approach fails, making both modes lossy for a large enough coupling between them. Our approach is to instead make the reflection-induced coupling between forward and backward modes vanishingly small, achieved by ensuring that the backward-propagating mode at the same energy has at least the opposite polarization, and potentially also the opposite angular momentum. \emph{Traversing a non-planar cavity in the opposite direction reverses the image rotation, breaking inversion symmetry of the system and thus making it helical: the two polarization states of the cavity are split out in frequency (see Appendix~\ref{appdx:pol_eigenstates_of_cav}), while forward- and backward-traveling modes of the same helicity remain degenerate~\cite{Jia2018}. This frequency splitting of the cavity polarization states is the primary feature enabling backreflection suppression.}

The total effect of backscattering can be described as a product of matrix elements in three sectors: (i) polarization; (ii) mode envelope; and (iii) transverse mode profile\footnote{The latter two could also be thought of as specific cases of a more general ``spatial mode profile'' sector.}. We discuss each of these sectors, finding that sectors (i) and (iii) benefit from non-planarity. 

\section{Polarization sector}

We first analyze backreflections in the polarization sector. The eigen-polarizations of a cavity can be found using Jones calculus~\cite{Jones1941} (see Appendix~\ref{appdx:pol_eigenstates_of_cav}). The cavity round-trip Jones matrix is composed of Jones matrices for individual optics arising from birefringence $B(\delta)$, where $\delta$ is the birefringent phase shift, as well as rotation $R(\theta)$ of the coordinate system into the local $s$ and $p$ polarization axes, where $\theta$ is the required rotation angle. For twisted cavities, the forward and backward round-trip Jones matrices need not be identical. These Jones matrices, $J_{\text{F}}$ and $J_{\text{B}}$, can be written as
\begin{eqnarray} \label{eq:JF_JB}
	\hspace{-0.05\textwidth} J_{\text{F}} = \prod_{j=1}^{n} B(\delta_{j}) R(\theta_{j}) && \hspace{0.05\textwidth}  J_{\text{B}} = \prod_{j=n}^{1} B(\delta_{j}) R(-\theta_{j-1})
\end{eqnarray}
\noindent
That is, the order of the index $j$ over the cavity optics is reversed for forward versus backward matrices and the rotation angles become negative (and $\theta_{0} \equiv \theta_{n}$). As shown in Appendix~\ref{appdx:JF_JB_same}, $J_{\text{F}}$ and $J_{\text{B}}$ have the same eigenvalues (and thus resonance energies); however, the eigenstates can differ.

\begin{figure*}
	\centering
	\includegraphics[width=0.98\textwidth]{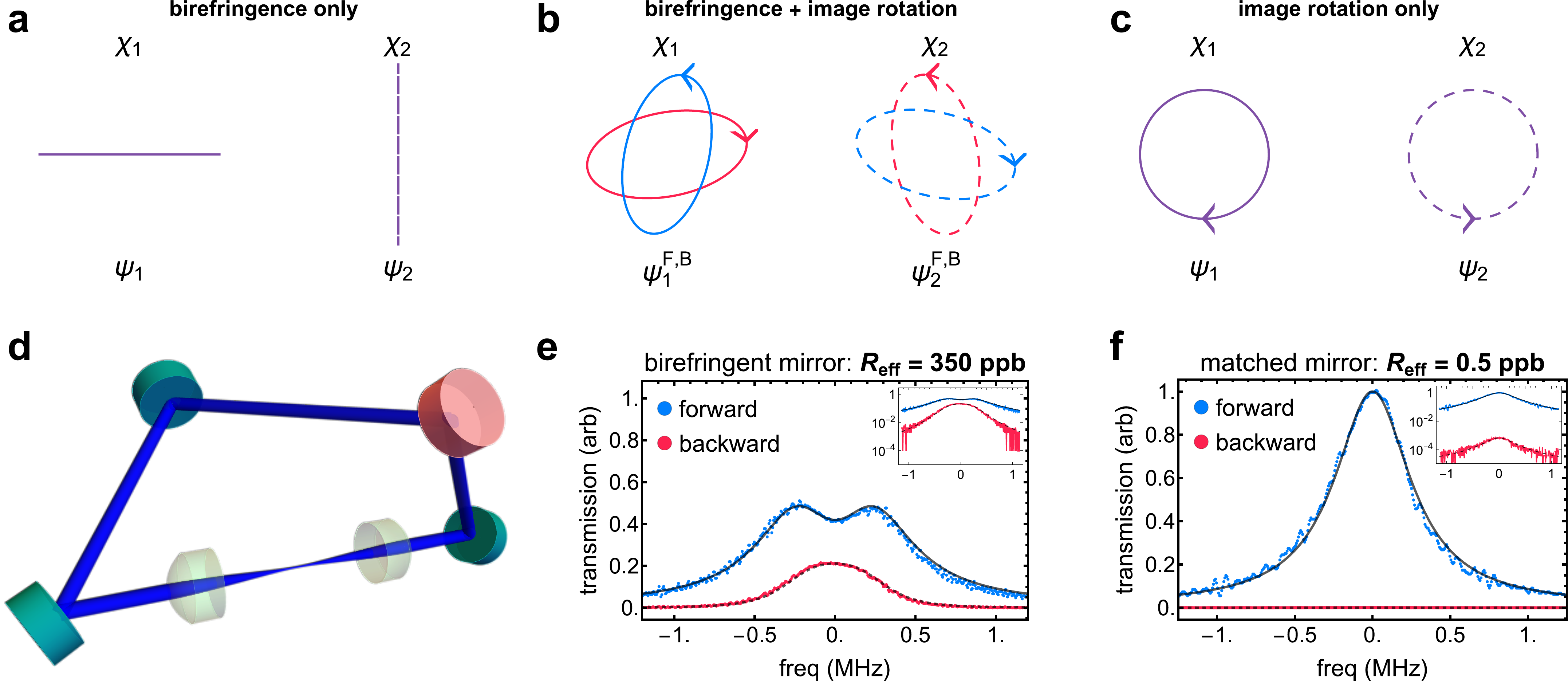}
	\caption{
	\textbf{Polarization suppression of backscattering} Cavity polarization eigenstates labeled by their eigenvalues $\chi_{i}$ and eigenvectors $\psi_{i}$ are shown in (a-c). (a)~A cavity with only birefringence has linearly polarized eigenstates (along the birefringent axes). (c)~A non-planar cavity with image rotation has circularly polarized eigenstates. (b)~In a cavity with both image rotation and birefringence, the modes are in general elliptically polarized. The forward (\Fclr{}) and backward (\Bclr{}) polarization modes are shown for each eigenvalue $\chi_{i}$, though they overlap exactly in (a) and (c). (d)~Cavity arrangement. The pink mirror is switched between a birefringent mirror and a non-birefringent mirror; see main text for details. Measured cavity transmission for the forward (\Fclr{}) and backward (\Bclr{}) modes are shown in (e)~birefringent and (f)~non-birefringent configurations. Simultaneous fits (black) to the forward and backward data extract the effective reflectivity $R_{\text{eff}}$. The same data is shown on a logarithmic scale in the insets. Much lower excitation of the backward mode is evident in the birefringence-canceled case.
	}
	\label{fig:pol_Jones_sector}
\end{figure*}

For the case of birefringent phase shifts arising from non-normal incidence on dielectrics in a planar resonator (see Fig.~\ref{fig:pol_Jones_sector}a), the forward and backward matrices are the same: $J_{\text{biref}}^{F} = J_{\text{biref}}^{B} = B(\delta) $. The resulting polarization eigenstates are linear, oriented along the birefringence axes. In a linear polarization basis, the Jones matrix for the reflection operator is $J_{\text{refl}}^{\text{lin}} = \begin{pmatrix}1&0\\0&-1\end{pmatrix}$. When a polarization eigenmode is backreflected, it is still resonant: reflection acting on an eigenvector of $J_{\text{biref}}^{F}$ results in an eigenvector of $J_{\text{biref}}^{B}$ with the same eigenvalue.

For the case of an image rotation only (see Fig.~\ref{fig:pol_Jones_sector}c), the backwards-traveling mode image-rotates in the opposite direction. However the $z$-direction inverts as well, so the two Jones matrices are again the same: $J_{\text{rot}}^{F} = J_{\text{rot}}^{B} = R(\Theta)$. The resulting eigen-polarizations are thus circular. Recall that the Jones basis vectors express polarization with respect to propagation direction; that is, they actually describe helicity. Reflection flips the helicity of light, so the Jones matrix for reflection in the circular polarization basis is $J_{\text{refl}}^{\text{circ}} = \begin{pmatrix}0&1\\1&0\end{pmatrix}$. Therefore, reflection acting on an eigenvector of $J_{\text{rot}}^{F}$ gives an eigenvector of $J_{\text{rot}}^{B}$ with the opposite eigenvalue. That is, for circular polarization states in a twisted cavity (whose Jones matrix is a rotation), \emph{backreflections are energetically forbidden.} Intuitively, this can be seen from the polarization frequency splitting arising from the circular polarization vector rotating either with or against the image rotation. When a reflection occurs, the image rotation changes sign, but the polarization rotation (i.e., spin angular momentum, not helicity) does not. 

To summarize, backreflection of linear polarization states is allowed: light resonant in the forward direction is also resonant in the backward direction. Circularly polarized eigenmodes of a twisted cavity, however, are protected against backreflections. Light of a given polarization mode finds no density of states to backscatter into.

In the general case  of both birefringence and rotations, the cavity polarization modes are elliptical (see Fig.~\ref{fig:pol_Jones_sector}c). For such a cavity, we can calculate the degree of backreflection suppression from the matrices $J_{\text{F}}$ and $J_{\text{B}}$. Let $\psi^{\text{F}}_{1,2}$ and $\psi^{\text{B}}_{1,2}$ be the eigenvectors of $J_{\text{F}}$ and $J_{\text{B}}$, with eigenvalues $\exp(i \chi_{1,2})$ ($J_{\text{F}}$ and $J_{\text{B}}$ have the same eigenvalues). The matrix element of the reflection operator is then given by
\begin{equation}
	\alpha_{\text{pol}} = \bra{\psi^{\text{B}}_{j}} J_{\text{refl}} \ket{\psi^{\text{F}}_{j}}
\end{equation}
\noindent
for $j=1,2$. The magnitude of this matrix element (squared) quantifies the degree of backreflection allowed in the polarization sector.

To demonstrate this effect, we measure backreflections in a cavity whose schematic is shown in Fig.~\ref{fig:pol_Jones_sector}d. We employ a non-planar cavity with two intracavity plano-convex lenses that act as reflectors. The lenses are super-polished\footnote{Polished and coated by Perkins Precision Developments.} for low surface roughness ($<2$~\AA~rms) and are AR-coated to $R_{\text{AR}} = 250$~ppm at 780~nm. The lower input/output mirrors are designed for zero birefringent phase shift between $s$ and $p$ polarization at their $28^{\circ}$ angle of incidence (AOI)\footnote{Designed and manufactured by Five Nine Optics.}. Minimizing this phase difference $\delta_{sp}$ is critical; we found that a coating with $\delta_{sp}=0\pm1^{\circ}$ specification performed notably better than a different coating with $\delta_{sp}=0\pm3^{\circ}$. The upper mirrors have a spatial rotation of $89^{\circ}$ between their respective $s$ and $p$ axes. Thus, if these two mirrors have the same coating, any residual birefringence nearly cancels after reflection off both mirrors, as $s$-polarization for the first mirror very nearly becomes $p$-polarization for the second mirror. We can vary the birefringence by using two upper mirrors from the same coating run (thus canceling birefringence), or by using mirrors from different coating runs, leaving a residual birefringent phase shift. The finesse of this cavity is $\mathcal{F} \approx 3900$, set primarily by the input/output mirror transmission (see Appendix~\ref{appdx:hi_F_lensCav}).

The results for backreflections of the fundamental cavity mode are shown in Fig~\ref{fig:pol_Jones_sector}. For the birefringent case (Fig.~\ref{fig:pol_Jones_sector}e), the forward mode begins to hybridize with the backward mode (cf. Fig.~\ref{fig:reflector_in_a_cavity}). Even in this case, the effective reflectivity ($R_{\text{eff}}^{\text{biref}} = 350$~ppb) is about 700 times lower than the bare reflectivity of the AR-coated lens ($R_{\text{AR}} = 250$~ppm). Furthermore, the above analysis assumes a single reflector, while in reality, this cavity has reflections at 4 lens surfaces. This $>700\times$ reduction in reflectivity from the free-space value primarily results from the round-trip image rotation still dominating over the un-canceled birefringence, meaning that the resulting polarization states are still nearly circular. From the measured birefringence of the mirrors, we only expect $|\alpha_{\text{pol}}|^{2}\sim6\%$ overlap between the forward and backward cavity polarization modes with the same eigenvalue. There is also another mechanism of backreflection suppression at work related to the spatial mode profile of the cavity mode (see Sec.~\ref{sec:spatial_mode_sector}).

For the non-birefringent case, we use mirrors designed for $\delta_{sp} = 0 \pm 10^{\circ}$ at the $45^{\circ}$ AOI of the upper mirrors. Since both upper mirrors come from the same coating run in this case, the $89^{\circ}$ spatial rotation between these mirrors cancels much of any residual birefringence of this coating. As shown in Fig.~\ref{fig:pol_Jones_sector}f, we see a \emph{further} 700-fold reduction in backreflections, plunging the effective reflectivity below 1 ppb. This is almost 1 million times lower reflectivity than the best-achievable AR-coatings in free space, and results in a negligibly-perturbed forward-propagating mode. Despite having four changes of refractive index per cavity round trip, the cavity polarization state is sufficiently circular that backreflections are almost entirely forbidden.


\section{Spatial mode sectors} \label{sec:spatial_mode_sector}

In addition to the polarization sector, the spatial mode profile matrix element between the initial and target state must be nonzero to allow backreflection to occur. Due to the image rotation and axial symmetry of the twisted cavity in Fig.~\ref{fig:pol_Jones_sector}d, the resulting eigenmodes must be invariant (up to a phase) under rotation and indeed may be shown to be Laguerre-Gauss (LG) modes ~\cite{schine2016synthetic}. The LG transverse mode profiles in a two-dimensional plane perpendicular to the propagation axis are given by 

\begin{align}
	u_{\ell,p}(r,\phi) =
			&\frac{c_{\ell, p}}{w} \left( \frac{r\sqrt{2}}{w} \right)^{|\ell|} \exp\left(-\frac{r^{2}}{w^{2}}\right)
			L_{p}^{|\ell|}\left(\frac{2 r^{2}}{w^{2}}\right) \nonumber \\
			&\times \exp\left( -i \frac{2\pi}{\lambda} \frac{r^{2}}{2 R} \right) \exp\left( i (2p + |\ell| + 1) \psi \right) \nonumber \\
			&\times \exp\left( -i \ell \phi \right) 
\end{align}

\noindent
where $r$ ($\phi$) is the radial (azimuthal) coordinate. The mode indices $\ell$ and $p$ are integers with $p \geq 0$, $L_{p}^{\ell}\left(x\right)$ are generalized Laguerre polynomials, and $\psi$ is the Gouy phase. The local beam waist $w$ and wavefront radius of curvature $R$ together define the complex beam parameter ($q$-parameter) as $\frac{1}{q} = \frac{1}{R} - i \frac{\lambda}{\pi w^{2}}$, where $\lambda$ is the wavelength of the light. The normalization constant $c_{\ell, p} = \sqrt{\frac{2p!}{\pi\left( p+|\ell| \right)!}}$ ensures that $\bra{	u_{\ell,p}} u_{\ell,p} \rangle = 1$. The three phase factors represent wavefront curvature, Gouy phase, and angular momentum in the form of a phase winding/optical vortex, respectively.

Consider how the cavity mode profiles transform under a reflection. The radial coordinate $r$ remains unchanged, while $\phi \rightarrow -\phi$. Since the sign of $\ell$ only matters in the phase winding term, the reflection thus has the effect of taking $\ell \rightarrow -\ell$. That is, the forward-propagating mode with angular momentum $\ell$ has the same phase profile as the backward-propagating mode with angular momentum $-\ell$. Furthermore, the backward-propagating mode at a given plane has the opposite sign wavefront curvature $R$ as the forward-propagating mode: if the forward mode has a diverging wavefront at a given plane, the backward mode is converging.

The matrix elements of reflection in the spatial profile sector between forward- and backward-traveling modes with the same mode indices are thus given by

\begin{align}
	\alpha_{\text{spatial}} &= \bra{u_{\ell, p}^{B}(z_{\text{refl}})} \hat{\mathcal{R}} \ket{u_{\ell, p}^{F}(z_{\text{refl}})} \nonumber \\
							&= \int r\,dr\,d\phi \,\, u_{\ell, p}^{q_{B}} (r, \phi)^{*} u_{-\ell, p}^{q_{\text{refl}}} (r, \phi) \label{eq:spatial_overlap_integral}
\end{align}

\noindent
where $\hat{\mathcal{R}}$ is the reflection operator, $z_{\text{refl}}$ is the plane of the reflector, and $F$ ($B$) indicates the forward (backward) mode. The $q$-parameter after reflection $q_{\text{refl}}$ can be found from action of the reflector's $ABCD$ matrix on the incident $q$-parameter $q_{F}$, and $q_{B} = -q_{F}^{*}$ is the $q$-parameter of the backward-traveling mode at $z_{\text{refl}}$.

From this expression, we find something noteworthy: regardless of $q_{F}$, \emph{all of these matrix elements for $\ell \neq 0$ are zero}, as the integral over the angle $\phi$ results in a Kronecker delta $\delta_{\ell, -\ell} = \delta_{\ell,0}$. We therefore analyze the $\ell \neq 0$ and the $\ell = 0$ cases separately. The integral Eq.~\ref{eq:spatial_overlap_integral} can be computed in closed form, even for the general case of a curved reflector and for all $\ell$. The result of the integral is:

\begin{equation} \label{eq:qmatch_closedform}
	\alpha_{\text{spatial}}^{\ell, p} = \delta_{\ell0}\sum_{j=0}^{p} \sum_{k=0}^{p} \gamma^{j+k+1} (-1)^{j+k} \begin{pmatrix}p\\j\end{pmatrix} \begin{pmatrix}p\\k\end{pmatrix} \frac{(j+k)!}{j!k!}
\end{equation}

\noindent
where $\gamma = -\frac{i \rho}{\zeta^{2} + \rho\left( \zeta - i \right) + 1}$, $\rho = \text{ROC}_{\text{refl}} / z_{\text{R}}$ is the reflector's radius of curvature in units of the Rayleigh range $z_{\text{R}}$, and $\zeta = z / z_{\text{R}}$ is the $z$-coordinate in units of $z_{\text{R}}$ (where $z=0$ is the waist).

\subsection{Mode envelope} \label{sec:q-matching}

\begin{figure*}
	\centering
	\includegraphics[width=0.95\textwidth]{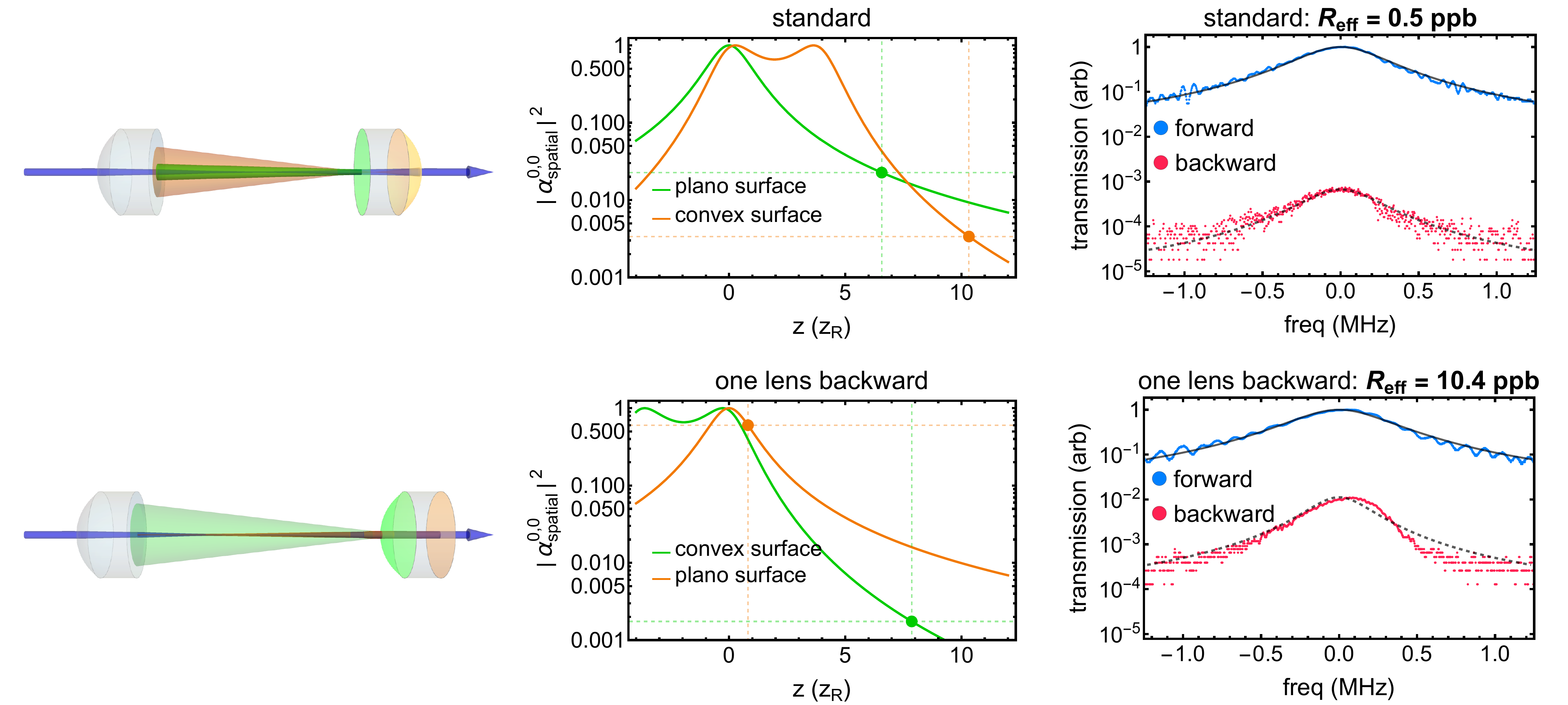}
	\caption{\textbf{Spatial mode-matching suppression of backscattering.} Standard (backward) orientation of the rightmost lens is shown in the top (bottom) row. Left column: lenses are shown, along with the cavity mode (blue; propagating left-to-right) and backreflected beam profiles from the first and second surfaces (green and orange, respectively) of the rightmost lens, in order of incidence. We ignore here reflections from the leftmost lens. The lenses have an outer diameter of 7.75 mm. The tubes indicating the mode size have radius equal to 1.92 times the local beam waist; that is, 99.94\% of the beam's power is contained within the tube, and losing all power outside of this tube would limit the finesse to $10^{4}$. Note that the reflection off of the second surface of the backwards lens is best mode-matched to the backward-traveling fundamental mode. Middle column: Eqn.~\ref{eq:qmatch_closedform} is plotted for the lens surfaces, showing the spatial overlap integral vs. position parameterization (see main text) for each reflecting surface. The reflecting surfaces in our cavities are indicated by the points along the curves. Note that the backwards-orientation plano surface has substantially higher $|\alpha_\text{spatial}^{0,0}|$ than any of the other surfaces. Right column: Experimental backreflection data. Due to the improved mode matching, the effective reflectivity coefficient $R_{\text{eff}}$ in the backwards-lens case is about 20$\times$ greater than in the standard configuration (top right is the same data as in Fig.~\ref{fig:pol_Jones_sector}f). The standard configuration is thus more favorable for reducing backscatter.
	}
	\label{fig:QmatchingFigure}
\end{figure*}

Equation~\ref{eq:qmatch_closedform} mainly quantifies the observation that a reflection should be well-mode-matched into the backwards traveling mode in order to occur. This is visualized in the left column of Fig.~\ref{fig:QmatchingFigure}. This mode-matching can be increased by turning one lens to be backwards, such that the cavity mode encounters the lens' flat surface when it is large and less-divergent. This gives better mode-matching of the reflection into the backward mode, and thus a 20$\times$ increase in the effective reflectivity $R_{\text{eff}}$ (right column of Fig.~\ref{fig:QmatchingFigure}).

For the standard lens orientation, we see from Fig.~\ref{fig:QmatchingFigure}b that $|\alpha_{\text{spatial}}|^{2} \sim 0.02$, which provides a factor of 50 suppression of backreflections compared to the free space value. This factor of 50 due to mode mismatch accounts for the rest of the $\sim700\times$ reduction of the free space reflectivity $R_{\text{AR}}$ to the birefringent value $R_{\text{eff}}^{\text{biref}}$ in Fig.~\ref{fig:pol_Jones_sector}f.

\subsection{Backreflections with angular momentum}

\begin{figure*}
	\centering
	\includegraphics[width=0.95\textwidth]{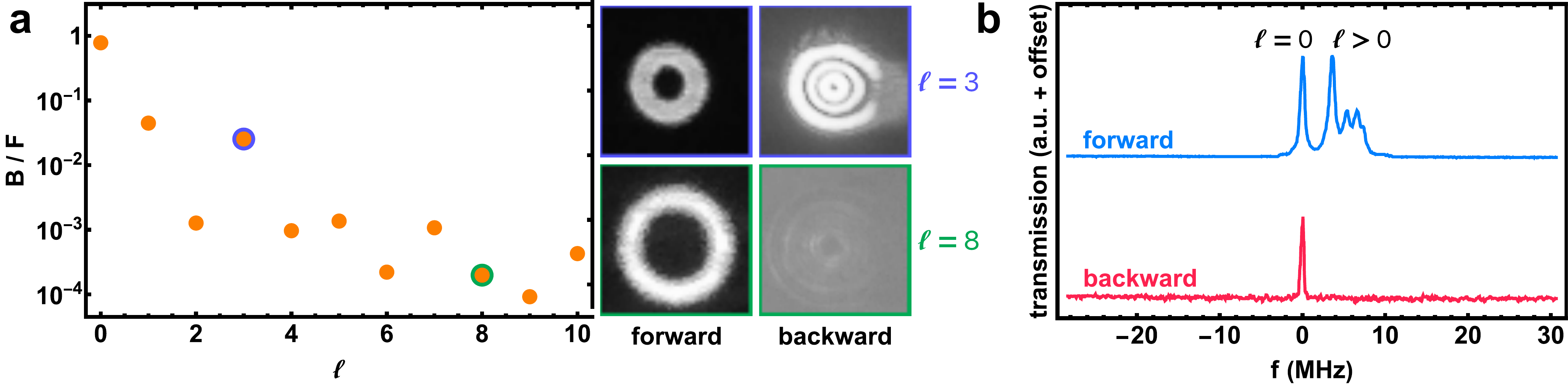}
	\caption{\textbf{Symmetry-protection from backscattering.} We demonstrate the additional backreflection suppression for beams carrying orbital angular momentum. In order to make the signal large enough to be observed, we increase backreflections by using a birefringent upper mirror, and placing one lens in the backward configuration.  (a) Backreflections vs orbital angular momentum $\ell$. Nonzero OAM provides one to four orders of magnitude backreflection suppression, limited entirely by backscatter into different modes. Example forward and backward modes illustrating this are shown at right. Evidently, for the lens spacing used there is an accidental degeneracy of the forward-traveling $(\ell=3,\,p=0)$ mode with a backward-traveling $p=2$ mode (presumably with $\ell = -3$). This resonant mode conversion~\cite{Stone2021} allows substantially greater backscatter than, e.g., the $\ell = 8$ case. The $\ell = 8$ ring mode scatters only very weakly into a high-$p$ mode, which is faintly visible on the backward path camera. Note that the image brightness scales are arbitrary relative to each other, as is the relative spatial scale between the forward and backward cameras (there is also some clipping visible in the backward $\ell=3$ image). (b) Setting the lens separation such that every $\ell \geq 0, \, p=0$ mode is nearly degenerate (this is special to our cavity design) shows that only the $l=0$ mode backreflects. Forward (backward) mode transmission shown in \Fclr{} (\Bclr{}).  The backreflection splits the $\ell=0$ mode out from the cluster of nearby, nearly degenerate $l>0$ modes, which do not backreflect. The vertical scales of the forward and backward traces are not equal, and are offset for clarity.
	}
	\label{fig:OAM_TR_backscatter}
\end{figure*}

We saw from the $\phi$-integral in Eq.~\ref{eq:spatial_overlap_integral} that LG modes with nonzero angular momentum cannot backreflect into themselves. To experimentally probe this, we increase cavity backreflections by using the birefringent mirror configuration of Fig.~\ref{fig:pol_Jones_sector}e, and reversing the orientation of one of the lenses (Fig.~\ref{fig:QmatchingFigure}, bottom row).

Using holographic beam-shaping with a digital micromirror device~\cite{Zupancic2016}, we inject a desired LG ring mode (i.e., $p=0$) into the cavity and observe its backreflection behavior. As shown in Fig.~\ref{fig:OAM_TR_backscatter}a, while $\ell=0$ shows substantial backscatter, we see backreflection suppression by one to four orders of magnitude for nonzero angular momentum. In fact, by observing both the forward and backward modes on CCD cameras, we find that this backreflection suppression is limited entirely by the availability of other modes to backscatter into. Since the $q_{\text{refl}}$ is different from $q_{B}$ in Eq.~\ref{eq:spatial_overlap_integral} (unless the reflector matches the phase front of the cavity mode), the forward and backward mode bases are not mutually orthonormal, so a mode $u^{F}_{\ell, p}$ could backscatter into a mode $u^{B}_{-\ell, p'}$ if there happens to be such an accidental near-degeneracy. This can be seen for $\ell=3$ in Fig~\ref{fig:OAM_TR_backscatter}a. Additionally, imperfections such as misalignment or optical surface wedge can introduce small couplings even between different $\ell$ states. Backreflection should be especially suppressed near intentional degeneracy points~\cite{schine2016synthetic,Kollar2015}, where modes cluster up, and accidental degeneracies are less likely.

In the lowest backreflection configuration (non-birefringent, standard lens orientation as in Fig.~\ref{fig:pol_Jones_sector}f), backreflections of higher-angular momentum states are unobservably small. The suppression quantified for $\ell>0$ modes in the higher-backreflection configuration (birefringent, one lens backward) of Fig.~\ref{fig:OAM_TR_backscatter} indicates that beams carrying orbital angular momentum experience effective reflectivities in the lowest backreflection configuration as low as sub-part-per-\emph{trillion}.

\section{Conclusions}
We have demonstrated the ability to incorporate optics inside of high-finesse optical cavities. Lenses and modulators could dramatically expand the capabilities of such resonators, as they do for free space optics. By engineering the polarization properties of the cavity eigenmodes, we suppress the effect of intracavity backreflections by six orders of magnitude, resulting in sub-part-per-billion effective reflectivities. Beams carrying orbital angular momentum provide even greater suppression due to their phase winding, limited only by the available density of states for mode conversion, and yielding effective reflectivities at the part-per-trillion level. We have explored the effects on backreflection of both polarization and spatial mode profile in these non-planar cavities.

Unlocking more of the optics toolkit for use in optical resonators could dramatically expand the possibilities for high-cooperativity interfacing with quantum emitters. Aspheric lenses can be used to shape cavity spectra~\cite{Jaffe2021}, and high-NA systems such as optical tweezers or atom imaging could be Purcell-enhanced. At high-NA, where paraxial approximations break down, new regimes such as strong light-matter coupling in the presence of vector optics effects like spin-orbit coupling~\cite{zeppenfeld2010calculating, Naidoo2016} may be explored. Incorporating non-planarity into standard, mirror-only high-finesse ring cavities may also reduce undesired backreflections~\cite{Krenz2007}.

\begin{acknowledgments}
This work was supported by AFOSR Grant FA9550-18-1-0317, and AFOSR MURIs  FA9550-19-1-0399 \& FA9550-16-1-0323.
\end{acknowledgments}

\appendix
\section{Polarization eigenstates of a cavity} \label{appdx:pol_eigenstates_of_cav}
For most cavities, polarization is an unimportant degree of freedom. The two polarization states typically have the same energies; a degeneracy limited only by stress-induced birefringence in the mirrors of very high finesse 2-mirror cavities~\cite{Hall2000}, or non-normal mirror incidence in planar running-wave cavities~\cite{kruse2003cold}. For the non-planar cavities we describe here however, there is special structure to the polarization eigenmodes.

Just as ray transfer ($ABCD$) matrices are used to calculate the spectra and eigenmodes of an optical resonator~\cite{siegman86}, Jones matrices~\cite{Jones1941} can be used to calculate the polarization eigenmodes of a cavity. The round trip Jones matrix of a cavity is calculated by composing the individual Jones matrices of optical elements. The two relevant types of Jones matrices here will be birefringence and spatial rotation.

For a linear phase retarder whose fast axis is aligned with the $x$-axis of the coordinate system, the matrix describing its action on a Jones vector describing a polarization state in the basis of horizontal ($\ket{H}$; $x$ direction) and vertical ($\ket{V}$; $y$ direction) polarization states takes a simple form:

\begin{equation}
	B(\delta) = \begin{pmatrix}
		e^{-i\frac{\delta}{2}} & 0 \\
		0 & e^{i\frac{\delta}{2}}
	\end{pmatrix}
\end{equation}
where $\delta$ is the phase retardation between the fast and slow axes. That is, the $\ket{H}$ and $\ket{V}$ basis states acquire a relative phase shift $\delta$.

For our non-planar cavities, we will rotate the local coordinate system between each optical element over a cavity round trip such that the coordinate $x$-axis indeed corresponds to one of the birefringent axes (the $p$-polarization). These rotation matrices are given by:

\begin{equation}
	R(\theta) = \begin{pmatrix} \cos\theta & -\sin\theta \\ \sin\theta & \cos\theta \end{pmatrix}
\end{equation}

For an $n$-optic non-planar cavity, the cavity Jones matrix is then given by
\begin{equation} \label{eq:J_rt}
	J_{\text{rt}} = \prod_{j=1}^{n} B(\delta_{j}) R(\theta_{j}) 
\end{equation}

\noindent
where each optic's birefringent phase shift is $\delta_{j}$, and $\theta_{j}$ is the angle required to rotate the coordinate system at optic $j$ into the coordinate system of optic $j+1$.

Consider a cavity with no rotations, but with a single birefringent phase shift of $\Delta$. This cavity has the round trip Jones matrix $J_{\text{biref}} = B(\Delta)$. The eigenvectors (and thus, polarization eigenstates) are then simply $\ket{H}$ and $\ket{V}$, and the eigenvalues are $\chi_{\pm} = \exp(\pm i\Delta/2)$. The round trip phases acquired by the two polarization states differ by $\Delta$, which sets the birefringent polarization mode splitting.

Next, consider a non-birefringent cavity whose Jones vector consists purely of a rotation of nonzero angle $\Theta$, $J_{\text{rot}} = R(\Theta)$. The polarization eigenstates are then given by
\begin{equation} \label{eq:Jrot_evecs}
	\psi_{\pm} = \frac{1}{\sqrt{2}} \begin{pmatrix} 1\\ \pm i\end{pmatrix}
\end{equation}

\noindent
which can be recognized as the left- and right-handed circular polarization states $\ket{L} \equiv \frac{1}{\sqrt{2}}\left( \ket{H} + i \ket{V} \right)$ and $\ket{R} \equiv \frac{1}{\sqrt{2}}\left( \ket{H} - i \ket{V} \right)$, respectively.

The eigenvalues are $\chi_{\pm} = \exp(\pm i\Theta)$. Thus, the round trip phases acquired by the two polarization states differ by $2\Theta$. This corresponds to the circular polarization rotating either with or against the image rotation represented by the rotation matrix. A splitting of these polarization modes arises even in the absence of birefringence.

Note also that each reflection from a cavity mirror also contributes a reflection Jones matrix

\begin{equation}
	J_{\text{refl}}^{HV} = \begin{pmatrix} 1 & 0 \\ 0 & -1 \end{pmatrix},
\end{equation}
\noindent
written in the ${\ket{H}, \ket{V}}$ basis. The minus sign can be interpreted in a number of ways. In the case of non-normal incidence, it is clear that the tangential axis is reversed under reflection. More generally, including the case of normal incidence, the ``polarization'' basis states of the Jones calculus are defined with respect to a propagation direction (i.e., they are actually helicity states). This can be seen most clearly in the circular basis, where left- or right-handedness of the rotating electric field is defined with respect to the propagation direction. Upon reflection, the rotation direction of the electric field vector does not change, but the propagation direction does; the helicity therefore changes sign. A reflection thus performs the transformation $\ket{L} \leftrightarrow \ket{R}$. Converting this transformation into the $\{ \ket{H}, \ket{V} \}$ basis gives us $J_{\text{refl}}$ above.

\section{High-finesse lens cavity} \label{appdx:hi_F_lensCav}

The lenses used for the cavities in this work were manufactured by Perkins Precision Developments. They are plano-convex lenses with 5 mm radius of curvature made from Corning 7980-UV-1D fused silica. Both surfaces were super-polished to ultra-low surface roughness, $<2$~\AA~rms, and AR-coated to $R_{\text{AR}} = 250$~ppm at 780~nm. We found that this ultra-low surface roughness was critical to achieving low loss. Similar lenses with higher surface roughness (not spec'd) led to substantially greater intracavity loss.

\begin{figure}
	\centering
	\includegraphics[width=0.45\textwidth]{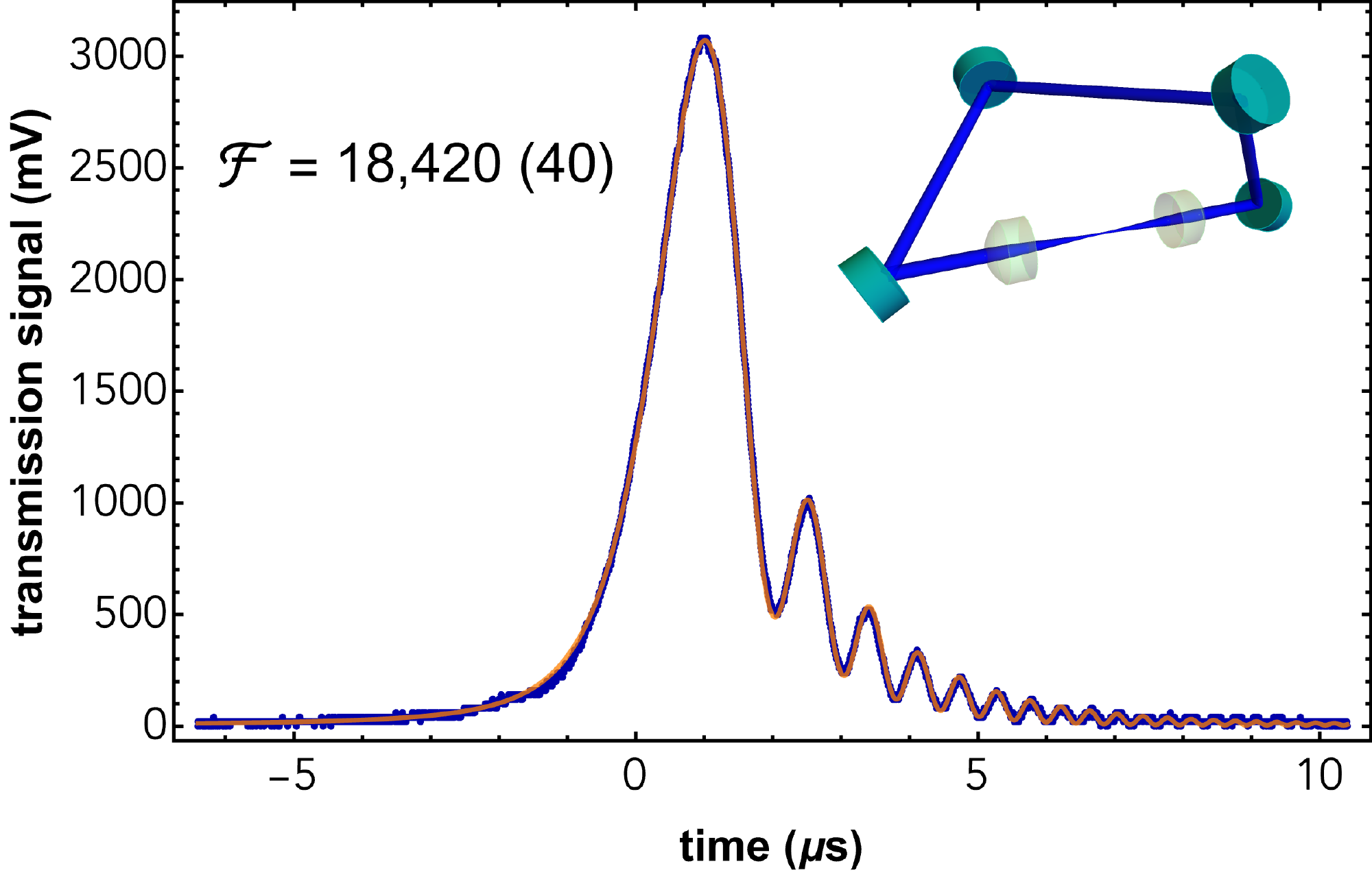}
	\caption{\textbf{High-finesse two-lens cavity.} This high-finesse cavity with two lenses is demonstrated as a proof of principle that high-quality intracavity optics are compatible with high finesse. Higher-transmission input/output mirrors were used for the rest of the work presented here, setting the finesse to $\mathcal{F}\approx3900$. 
	}
	\label{fig:hi_F_LensCav_Fig}
\end{figure}

As a demonstration of the cavity finesse possible with such lenses, we show here a twisted cavity where the 4 mirrors used supported a cavity finesse of $34\times 10^{3}$. After inserting two of these superpolished lenses into the cavity, the resulting finesse was $18.4\times 10^{3}$. A cavity ringdown~\cite{Poirson1997} from which this value was obtained is shown in Fig.~\ref{fig:hi_F_LensCav_Fig}. This corresponds to a loss of about 80~ppm per lens per round trip (note that each lens contributes two changes of refractive index per round trip).

\section{$J_{F}$ and $J_{B}$ share eigenvalues} \label{appdx:JF_JB_same}
In this appendix, we prove that $J_{F}$ and $J_{B}$, as defined in Eq.~\ref{eq:JF_JB} have the same eigenvalues. Let $\sigma(\mathbf{M})$ denote the spectrum, or set of eigenvalues, of a matrix $\mathbf{M}$. This spectrum is invariant under cyclic permutations
\begin{equation}
	\sigma(\mathbf{A}_{n}\mathbf{A}_{n-1}...\mathbf{A}_{2}\mathbf{A}_{1}) = \sigma(\mathbf{A}_{n-1}...\mathbf{A}_{2}\mathbf{A}_{1}\mathbf{A}_{n})
\end{equation}
as well as the transpose
\begin{equation}
	\sigma(\mathbf{M}) = \sigma(\mathbf{M}^{\intercal})
\end{equation}
We also note that the transpose of a product of matrices can be written as the reversed-order product of transposes:
\begin{equation}
	\left(\mathbf{A}\mathbf{B}\mathbf{C}\mathbf{D}\right)^{\intercal} = \mathbf{D}^{\intercal}\mathbf{C}^{\intercal}\mathbf{B}^{\intercal}\mathbf{A}^{\intercal}
\end{equation}

Taking the definitions of $J_{F}$ and $J_{B}$ from Eq.~\ref{eq:JF_JB}, we note that $J_{F}$ can be obtained from $J_{B}$ by transposing and cyclically permuting. By transposing, we get:
\begin{align}
	J_{B}^{\intercal} &= \left(\prod_{j=n}^{1} B(\delta_{j}) R(-\theta_{j-1})\right)^{\intercal} \nonumber \\
					  &= \prod_{j=1}^{n} R(-\theta_{j-1})^{\intercal}B(\delta_{j})^{\intercal} \nonumber \\
					  &= \prod_{j=1}^{n} R(\theta_{j-1})B(\delta_{j})
\end{align}
\noindent
where in the last line we used that $B(\delta_{j})$ is diagonal and that $R(\theta_{j})^{\intercal} = R(-\theta_{j})$. One cyclic permutation, moving $R(\theta_{0})\equiv R(\theta_{n})$ from the first to the last position then gives us $J_{F}$. This then allows us to write

\begin{equation}
	\sigma(J_{B}) = \sigma(J_{B}^{\intercal}) = \sigma(J_{F})
\end{equation}
\noindent
which tells us that $J_{F}$ and $J_{B}$ have the same eigenvalues.

\newpage
\bibliographystyle{naturemag}
\bibliography{backscatter_bibliography}

\end{document}